# OAM spatial demultiplexing by diffraction-based noiseless mode conversion with axicon

Junsu Kim [1], Jiyeon Baek [2], Hyunchae Chun[*], Seung Ryong Park[*]

*Abstract* — The physics of orbital angular momentum (OAM) carrying light has been well-defined since the 1990s. Leveraging its physical phenomena has become a significant focus in various areas of research. For instance, OAM is applied in hybrid free-space optical communication channels, like wavelength division multiplexing. Due to its orthogonality, OAM can be multiplexed and demultiplexed, enabling the use of multiple orthogonal OAM beams to achieve high-capacity optical communication systems. In this appliance, there will be instability of light because its phase distribution is quite complicated. This study focuses on spatial demultiplexing using Computer Generated Holography (CGH) based OAM diffraction. Earlier research on OAM diffraction primarily involved fork gratings. Spatial demultiplexing has phase distribution instability and noise problems in free space optics. Axicons have been studied extensively in the context of perfect vortex generation. Drawing inspiration from these two areas of physics, we propose a noiseless topological charge conversion through axicon properties in spatial demultiplexing using CGH. Furthermore, we estimate the axicon optical approximate solution that is appropriate for research purposes.

## I. INTRODUCTION

Orbital angular momentum (OAM) of light can be clarified through wave function formula expanded as the Laguerre-Gaussian (LG) mode [1]. It is also referred to as an optical vortex or a twisted photon because its wave front is shaped like a vortex. The LG mode is derived from the paraxial Helmholtz equation in cylindrical coordinates and the Laguerre polynomial $L_n^l$. In the LG mode wave function, the phase factor of the vortex rotation mode is represented by a special exponential term $e^{il\varphi}$ [2-7]. This $l$ denotes the OAM Topological charge (TC), which gives rise to the azimuthal phase distribution. The LG mode carrying the OAM state is estimated by TC.

Prior studies on the OAM application process consist of critical issues, such as interference and phase distribution instabilities, which remain significant challenges [8-13]. OAM in free-space optics is widely used in photonics and communication technologies due to its orthogonality and optical characteristics [14-16]. To achieve high optical resolution and clarity in high-capacity communication systems, OAM multiplexing and demultiplexing processes are indispensable. For instance, clarity in demultiplexing hybrid channel systems is of paramount importance, although modulating these systems is challenging because optical vortices originate from wave front with phase differences between individual waves. Accuracy in the Topological Charge Conversion (TCC) process is another major challenge for practical applications [17].

TCC is a fundamental process these days, and its phase distribution control determines its value. A significant approach to this was made with Dammann grating OAM diffraction [18]. It was successfully applied in variable ways for the mux-demux process. We address these challenges in an unconventional way. We also aimed to construct a spatial demultiplexing system using a Spatial Light Modulator (SLM) based Computer Generated Holography (CGH). In this work, complex gratings surface with the SLM CGH has been used for OAM diffraction [19-22].

Here, we propose a noiseless TCC using the TC transformation rule and Perfect Vortex (PV) principle [23-25]. The TC transformation occurs through the diffraction of the OAM beam by a fork grating. A fork grating is the superposition of a blazed grating and a helical phase pattern. The TCC of light is governed by the TC transformation rule. Light diffracted at a specific angle carries a diffraction order-based OAM light TC state. This property induces TC = 0, allowing for the detection of spatial demultiplexing with phase distribution in the experiment. A demultiplexing system could be built with this, based on diffraction order and TCC. However, the TCC process usually generates a lot of noise, which can be a problem for demultiplexing.

To resolve this problem, PV principle with an axicon to the SLM CGH grating superposition [26][27]. We show that the plane of the Depth of Focus (DOF) of the axicon can be utilized as the beam detecting plane for demultiplexing specific OAM light.

The Depth of Focus (DOF) optical properties of an axicon have been studied, which serves its optical purpose. We derive the TCC efficiency maximization equation from the optical properties of an axicon and fit it with experimental data. In the CGH, grating superposition induces complexity and diffraction development based on the TC transformation rule and the axicon. The process is not a form of superposition but rather a TC reconversion process. For this system, TC=1 mode can be converted to TC=0 mode by following the TC conversion rules based on the diffraction order. It can be described as a phase inverse transformation of the phase distribution for demultiplexing [28]. It is occasionally seen in various experiments with different helical phase manufacturing methods [29-31].

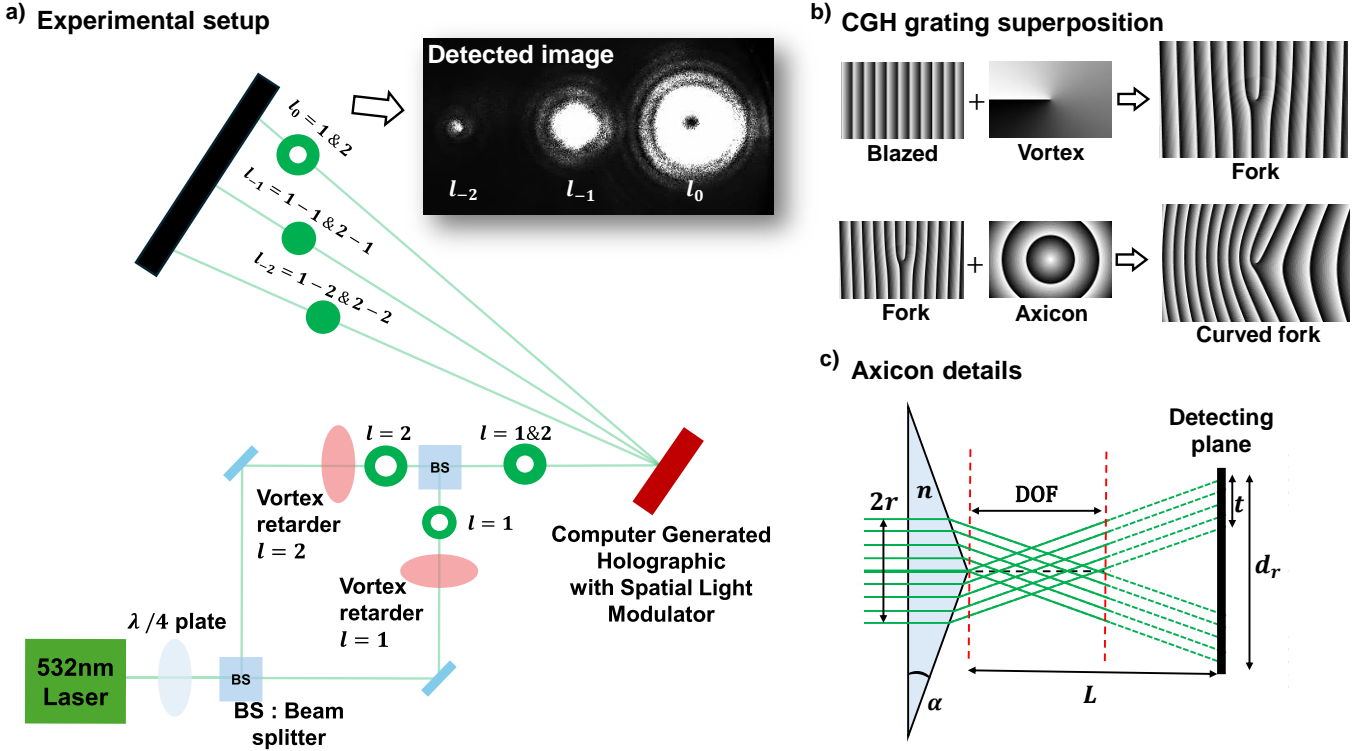

**Fig. 1.** (a) shows the experimental setup used in this study. The OAM beams with TC=1 and TC=2 is multiplexed by a combination of two beam splitters and two vortex retarders. The multiplexed beam is then demultiplexed using SLM CGH. The detailed is described in the text. The inset shows representative data obtained with a CMOS 2D image sensor in the experimental setup. (b) The fabrication process of the SLM CGH curved fork grating image used in this study is shown. The curved fork grating is a combination of a fork grating and an axicon. (c) The optical path of light through the axicon is shown. The apex angle $\alpha$ and the refractive index $n$ determine the optical path. The axicon can also be implemented using SLM CGH in the way shown in (c), and the parameters of angle $\alpha$ and refractive index $n$ can be adjusted by depth and gap length.

## II. METHOD & EXPERIMENT

In this experiment, a laser is used to generate 532 nm linearly polarized light. The laser power is 20 mW. The circularly polarized light, through a quarter wave plate is multiplexed with a combination of two beam splitters and two vortex retarders to create the OAM light with each modes TC = 1 and TC = 2. This light has been reflected onto the SLM CGH. The image of the reflected light is then captured using a CMOS 2D image sensor. The CMOS 2D image sensor has a resolution of 1440 × 1080 pixels and an imaging area of 4.968 mm × 3.726 mm.

The SLM CGH diffraction process is based on the TC transformation rule. The $l_{diffraction\ order}$ in Fig.1(a) indicates which OAM modes are included in each diffraction order. In this experiment, OAM light with TC = 1 and TC = 2 is superimposed, resulting in light diffraction. TC = 1 and TC = 0 is detected in $l_{-1}$, and TC = 0 and TC = -1 in $l_{-2}$. By identifying the diffraction order that includes TC = 0 in the final beam, one can determine the TC mode of the initial beam. The inset of Fig. 1(a) shows a demultiplexed image of $l_0$, $l_{-1}$ and $l_{-2}$ taken simultaneously. Measurements are taken zoomed in with a lens for precision.

Fig.1(b) shows the SLM CGH image that is employed. It features a curved fork grating, which is a combination of a fork grating and axicon. The fork grating consists of a blazed grating and a vortex plate. The groove density of the Blazed grating is 15.564 mm$^{-1}$, and its angle has been chosen to maximize the intensity of the $l_{-1}$ according to the purpose of the experiment. According to the TCC rule, the vortex helical phase mode determines TC modes for each diffraction order it contains. Thus, we can design appropriate demultiplexing processes for the initial TC state of the beam in the experiment.

The axicon acts as a spatial parity inversion for the detecting plane in Fig. 1(c) and plays a crucial role in our noiseless mode conversion method. An axicon is a type of conical optical lens that can be reproduced using CGH in the experiment. This has a special application in the phase distribution of the OAM beam [32][33]. While previous studies on axicon applications typically employed systems where the detecting plane is located beyond the DOF (DOF < L), our setup uses a detecting plane that coincides with the DOF (DOF = L). Here, L represents the detecting plane distance variable. This configuration results in a mirror inversion transformation of the beam's phase distribution with respect to the radial axis at the center. In brief, as light passes through the axicon depicted in Fig. 1(c), the inner portion of the beam is directed outward while the outer portion is directed inward. This principle forms the basis for the realization of noiseless TCC. Further details can be found in the part Ⅲ. The axicon is designed using two optical variables: apex angle ($\alpha$) and refractive index ($n$). In axicon CGH, $\alpha$ and $n$ can be adjusted by setting the depth and gap length. In the experiments, we quantitatively measured the TCC efficiency by varying the two variables that determine the shape of the axicon.

## III. RESULT & DISCUSSION

Fig. 2 shows only the $l_{-1}$ region of the total data obtained using the CMOS 2D image sensor. $\alpha$ and $n$ are 0.3 and 0.7, respectively. Fig.2(c) This image is a multiplexed beam of TC = 0 and OAM light TC = 1. This light appears at the $l_{-1}$ region when the TC = 1 and TC = 2 multiplexed light is diffracted by the curved fork grating. The final TC = 0 originates from the initial TC = 1 incident light, and the final TC = 1 originates from initial TC = 2 incident light. Fig. 2(a) and (b) show images of the $l_{-1}$ region captured by a CMOS 2D image sensor when only TC = 1 or TC = 2 is illuminated independently on the SLM CGH. Hence, detecting TC = 0 corresponds to the initial TC = 1 signal, and detecting TC = 1 corresponds to the initial TC = 2 signal.

TC = 0 shows the maximum light intensity at the center as seen in Fig. 2(a). TC = 1 results in a ring-shaped image with a minimum light intensity at the center, as shown in Fig. 2(b). In all of Fig. 2 (a), (b), and (c), the unstable part spreads outward in the Bessel beam mode [34]. Therefore, by extracting only the central light when TC = 0 and TC = 1 is multiplexed, TC = 0 light is detected. This means that the signal of TC=1 light is extracted without TC = 2 mixed in the incident light on the SLM CGH. We aim to spatially demultiplex light with multiple TC modes using this method. In this process, applying spatial parity inversion via an axicon enhances the clarity and purity of the valid TC mode during conversion.

To quantitatively measure the efficiency of the proposed axicon-based demultiplexing method, we define a metric called "TCC efficiency". The effective measurement area, indicated by the red dotted circle in the images of Fig.2, is defined as the light intensity $I_{Topolocial\ Charge}$, which is located at the center of each TC mode. In the experiment, the measurements are superimposed and have a total of three results for each TC = 0 and TC = 1. $I_0$ represents the light intensity within the effective measurement area for TC = 0, while $I_1$ represents it for TC = 1.

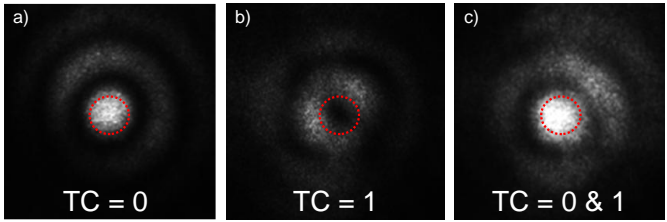

**Fig. 2.** Gray scale image of the result of diffraction by SLM CGH in CMOS with $n = 0.7$ for $\alpha = 0.3$, detected at $l_{-1}$. (a) When a beam with initial TC=1 is diffracted alone, a final TC=0 is measured. (b) final TC=1 is measured when the beam with initial TC=2 is diffracted alone. (c) final TC=0 and TC=1 is superimposed when the beam with initial TC=1 and TC=2 is diffracted.

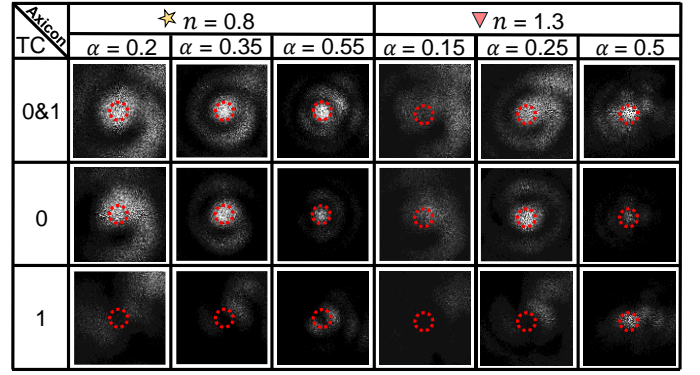

**Fig. 3.** Raw data from an experiment where the state is measured by varying the axicon parameters. The TC=0 effective measurement area for demultiplexing is the red dotted circle. These are three examples of results measured with fixed $n = 0.8$ and varying $\alpha = 0.35$ is the TCC efficiency maximization point. Three examples of results measured with fixed $n = 1.3$ and varying $\alpha = 0.25$ is the TCC efficiency maximization point. The TCC efficiency data fitting for this can be seen in Fig.4.

Measurements of individual TC modes are performed independently by separating beams with different TCs. Axicon's $\alpha$ and $n$ are measured as variables for each mode. The larger $I_0$ and the $I_1$ is closer to 0, it defined the higher TCC efficiency. $I_{total}$ represents the intensity of the two modes superimposed state, and the TCC efficiency is defined as $\frac{I_0-I_1}{I_{total}} \times 100$ by comparing the difference between the proportion occupied by TC = 0 and TC = 1. TCC efficiency indicates the degree of spatial demultiplexing and mode conversion. The experiment is conducted by varying α while keeping $n$ fixed and are repeated for several values of $n$. Fig. 4 shows the data points and the results of Gaussian fitting. For a fixed value of $n$, the TCC efficiency follows a Gaussian distribution as $\alpha$ varies. This suggests that the expectations and variance exist and can be analyzed. Based on the results, we infer a specific relationship between TCC efficiency and axicon variables. Looking at the respective peak positions, we observe that for $n > 1$, the $\alpha$ value at which TCC efficiency is maximized decreases as $n$ increases, whereas for $n < 1$. In addition, the variance of the Gaussian function exhibits a specific trend. Based on this relationship, we consider the function that maximizes TCC efficiency for a given $\alpha$ and $n$. This function can be derived through physical analysis. The optical properties of the axicon with respect to the detecting plane in Fig.1(c) has been used, which is a significant condition of the experiment. The axicon optical path is represented by the variables $r, L, d_r$ and DOF as shown in Fig.1(c). $r$ represents the initial beam radius and $d_r$ represents the final beam radius. The DOF and $d_r$ of the axicon are defined by Equation (1) and (2) below and are used to control the phase distribution. This provides spatial parity inversion to maximize TCC efficiency in demultiplexing. The specific optical conditions are DOF = $L$ and $d_r = 2r$, and the relationship between the equations satisfying these conditions can be summarized in terms of the two variables $\alpha$ and $n$ as follows.

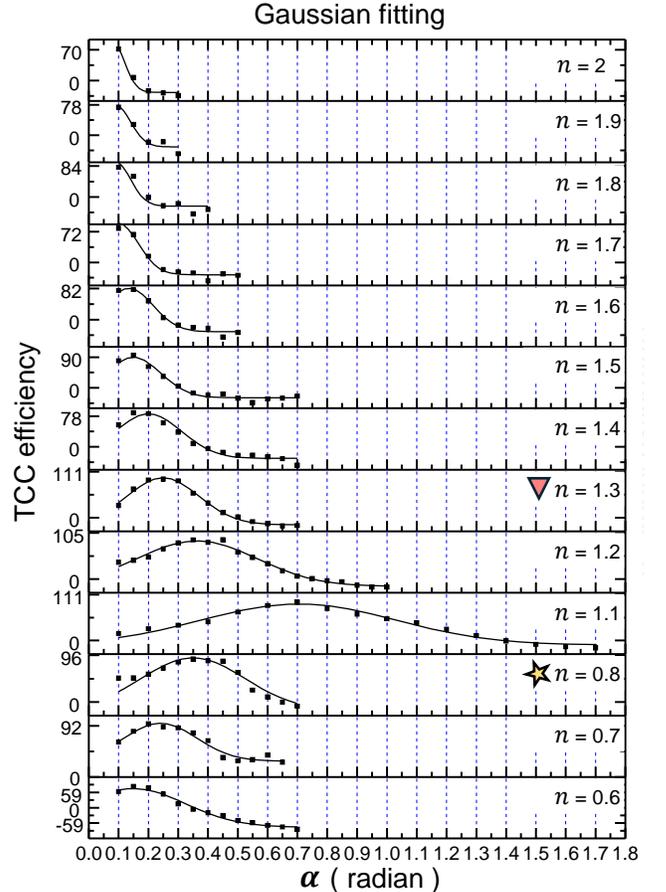

**Fig. 4.** The Gaussian fitting corresponding to the experimental results and the TCC efficiency is determined for $n$ and $\alpha$. For an explanation of TCC efficiency, please refer to the text above. TCC efficiency could be over 100 and under 0 because of the interference. Square dots are data points and solid line is Gaussian fitting.

$$DOF = \frac{r(1-n^2\sin^2\alpha)^{\frac{1}{2}}}{\sin\alpha\cos\alpha\,(n\cos\alpha - (1-n^2\sin^2\alpha)^{\frac{1}{2}})} = r \times A \quad (1)$$

$$d_r = \frac{2L\sin\alpha\left(n\cos\alpha - (1-n^2\sin^2\alpha)^{\frac{1}{2}}\right)}{n\sin^2\alpha + \cos\alpha\,(1-n^2\sin^2\alpha)^{\frac{1}{2}}} = 2L \times B \quad (2)$$

Under these additional conditions' the $\alpha$ is small and $n \neq 1$. We group the portion that depends on the variables $n$ and $\alpha$, substitute these parts with functions $A$ and $B$. Afterward, and apply the trigonometric limit condition to derive an approximation, which we then simplify.

$$A \approx \frac{1}{(n-1)\alpha} \,\&\, B \approx \tan[(n-1)\alpha] \rightarrow A \times B = 1 \quad (3)$$

$$\therefore \tan[(n-1)\alpha] = (n-1)\alpha \quad (4)$$

We derive an approximate solution (4) for the DOF point at which spatial parity inversion occurs [35][36]. The solution has been unified by defining a range of $\alpha$ and $n$ according to limitations of the axicon's optical structure and $\tan z = z$ equation [37]. The solution is applied to demultiplexing OAM for noiseless TCC and achieve the highest conversion efficiency. The fitted peaks (red dots in Fig. 5) match well with the solution graph (black solid line in Fig.5).

An additional interpretation is that the sigma value of the Gaussian fitting tends to decrease as $n$ increases when $n > 1$, while the slope of the function increases as $n$ increases. The sigma value is correlated with the corresponding gradient of graph. The differential equation for this solution can also be expressed in terms of the $\alpha$ and $n$ [38-39]. It is predicted that this equation could explain the sigma value in the Gaussian fitting. The experimental reliability of the Gaussian fitting peak positions depends on how well the differential equation and sigma fit. Here sigma(x2) (red solid line) and differential calculations (black solid line) are considerably correlated.

As shown in Fig. 5, The close match between the optical approximation function and the peak position fitting reinforces the reliability of our theoretical model. Those correspond to the TCC efficiency maximum points for specific axicon variables. This result shows that the axicon optical properties proposed in the solution are suitable for noiseless TCC in spatial demultiplexing for free-space optics. Additionally, the sigma value is related to the derivative of the solution.

However, under the condition $\alpha < 0.3$, sigma values for $n < 1$ tend to be larger than those for $n > 1$. Theoretically $|d\alpha/dn|$ has symmetry with $n = 1$ centered. This phenomenon cannot be fully explained by the obtained solution alone. We expect that the differences arise from the optical resolution related to the $n$. The $n > 1$ condition corresponds to a normal horn shaped axicon, whereas the $n < 1$ condition corresponds to an inverse shaped axicon. Furthermore, extreme conditions such as $n \rightarrow 1$ or $\pm\infty$ and $\alpha \rightarrow 0$ or $\pi/2$ could be measured if the experimental setup allows. Whether the solution holds under these extreme conditions remains an open question for future research. We focused on TC=1 demultiplexing, but this method has the same scalability and can be applied to all TC modes.

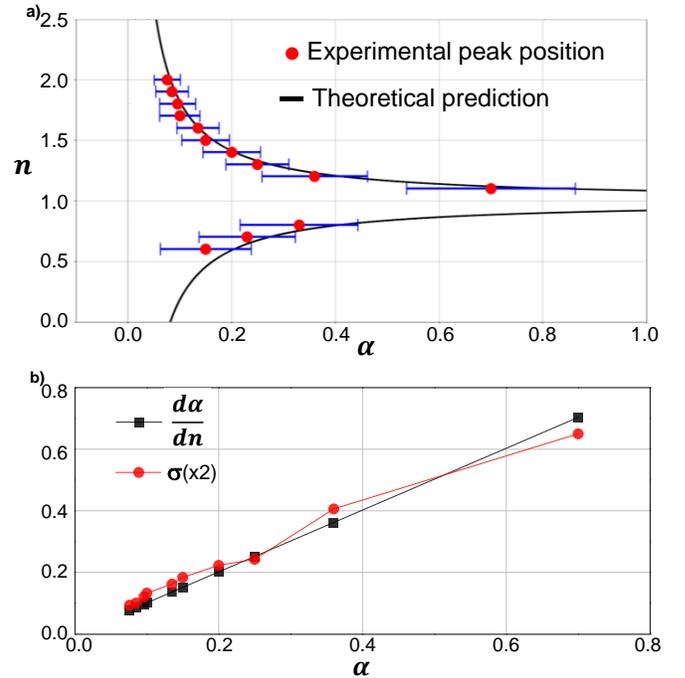

**Fig. 5.** Approximate solution we brought out from axicon optical properties. (a) $\tan[(n-1)\alpha] = (n-1)\alpha$ function graph as black solid line and Gaussian fitting peak position as red dots. Blue lines are sigma $\sigma$ of each Gaussian fitting. (b) $\frac{d\alpha}{dn} = \alpha\sec^2[(n-1)\alpha]$ calculation plot of differential equation and data plot of sigma $\sigma$(x2) of Gaussian fitting.

## IV. CONCLUSION

We propose both an experimental and theoretical approach to investigate the properties of axicon-applied gratings for noiseless TCC spatial demultiplexing. The study effectively demonstrated an axicon-applied SLM CGH grating for noiseless TCC, offering an improvement over conventional methods prone to noise interference. This achievement aligns with the optical approximation we derived, in which spatial parity inversion occurs at the DOF points of the axicon. The consistency between the experimental results and theoretical predictions highlights the robustness of the proposed method.

These results can be utilized to develop OAM diffraction-based TCC spatial demultiplexing systems with high precision and efficiency. This approach serves as a foundation for advancing OAM mux-demux systems, particularly in optical applications requiring high reliability and minimal noise interference. In addition, this principle facilitates the exploration of specific optical application of the axicon. It applies to systems that require phase distribution control and spatial parity inversion across various photonics applications. This thesis provides a comprehensive framework for understanding and utilizing axicon-based spatial demultiplexing technology.

This work is generally applicable to systems utilizing OAM. In the field of free-space optical communication, this noiseless spatial demultiplexing method enables high purity signal transmission, which is crucial for improving system performance. An additional advantage of this approach is that the demultiplexing process occurs only in a small region of a thin CGH. When applied to a demultiplexing detection device, this enables a more compact structure compared to other systems. CGH technology is also generally well defined for meta-surface studies and can be fabricated as ultrathin films using grating superposition. For diffraction methods such as Dammann gratings, this axicon appliance has the scalability to be easily applied by superposition. This study could be an extension for future advancements in OAM-based photonic systems and various applications.

## REFERENCES

[1] L. Allen, M. W. Beijersbergen, R. J. C. Spreeuw, and J. P. Woerdman, "Orbital angular momentum of light and the transformation of Laguerre-Gaussian laser modes," *Physical Review A*, vol. 45, no. 11. American Physical Society (APS), pp. 8185–8189, 01-Jun-1992.
[2] J. Leach, M. J. Padgett, S. M. Barnett, S. Franke-Arnold, and J. Courtial, "Measuring the Orbital Angular Momentum of a Single Photon," *Physical Review Letters*, vol. 88, no. 25. American Physical Society (APS), 05-Jun-2002.
[3] J. C. Neu, "Vortices in complex scalar fields," *Physica D: Nonlinear Phenomena*, vol. 43, no. 2–3. Elsevier BV, pp. 385–406, Jul-1990.
[4] J. C. Neu, "Vortex dynamics of the nonlinear wave equation," *Physica D: Nonlinear Phenomena*, vol. 43, no. 2–3. Elsevier BV, pp. 407–420, Jul-1990.
[5] S. M. Baumann, D. M. Kalb, L. H. MacMillan, and E. J. Galvez, "Propagation dynamics of optical vortices due to Gouy phase," *Optics Express*, vol. 17, no. 12. Optica Publishing Group, p. 9818, 27-May-2009.
[6] J. H. Poynting, "The wave motion of a revolving shaft, and a suggestion as to the angular momentum in a beam of circularly polarized light," Proc. R. Soc. Lond., A Contain. Pap. Math. Phys. Character **82**(557), 560–567 1909.
[7] M. Krenn and A. Zeilinger, "On small beams with large topological charge: II. Photons, electrons and gravitational waves," *New Journal of Physics*, vol. 20, no. 6. IOP Publishing, p. 063006, 07-Jun-2018.
[8] J. Ye *et al.*, "Multiplexed orbital angular momentum beams demultiplexing using hybrid optical-electronic convolutional neural network," *Communications Physics*, vol. 7, no. 1. Springer Science and Business Media LLC, 22-Mar-2024.
[9] Stoyanov, L., Topuzoski, S., Paulus, G.G. *et al.* Optical vortices in brief: introduction for experimentalists. *Eur. Phys. J. Plus* **138**, 702 (2023).
[10] J. Vickers, M. Burch, R. Vyas, and S. Singh, "Phase and interference properties of optical vortex beams," *Journal of the Optical Society of America A*, vol. 25, no. 3. Optica Publishing Group, p. 823, 27-Feb-2008.
[11] Manavalan, G., Balasubramaniam, G. M., & Arnon, S. (2024). Improving OAM-Based Optical Wireless Communication in Turbulence Using Conjugate Light Field and Hybrid Neural Networks. *Journal of Lightwave Technology*. Institute of Electrical and Electronics Engineers (IEEE). https://doi.org/10.1109/jlt.2024.3517712.
[12] A. Trichili, K. -H. Park, M. Zghal, B. S. Ooi and M. -S. Alouini, "Communicating Using Spatial Mode Multiplexing: Potentials, Challenges, and Perspectives," in *IEEE Communications Surveys & Tutorials*, vol. 21, no. 4, pp. 3175-3203, Fourthquarter 2019.
[13] S. Zheng and J. Wang, "Measuring Orbital Angular Momentum (OAM) States of Vortex Beams with Annular Gratings," *Scientific Reports*, vol. 7, no. 1. Springer Science and Business Media LLC, 17-Jan-2017.
[14] H. Huang *et al.*, "100 Tbit/s free-space data link enabled by three-dimensional multiplexing of orbital angular momentum, polarization, and wavelength," *Optics Letters*, vol. 39, no. 2. Optica Publishing Group, p. 197, 03-Jan-2014.
[15] J. C. García-Escartín and P. Chamorro-Posada, "Quantum multiplexing with the orbital angular momentum of light," *Physical Review A*, vol. 78, no. 6. American Physical Society (APS), 11-Dec-2008.
[16] E. N. Leith and J. Upatnieks, "Reconstructed Wavefronts and Communication Theory*," *Journal of the Optical Society of America*, vol. 52, no. 10. Optica Publishing Group, p. 1123, 01-Oct-1962.
[17] A. Longman and R. Fedosejevs, "Mode conversion efficiency to Laguerre-Gaussian OAM modes using spiral phase optics," *Optics Express*, vol. 25, no. 15. Optica Publishing Group, p. 17382, 11-Jul-2017.
[18] T. Lei *et al.*, "Massive individual orbital angular momentum channels for multiplexing enabled by Dammann gratings," *Light: Science & Applications*, vol. 4, no. 3. Springer Science and Business Media LLC, pp. e257–e257, 13-Mar-2015.
[19] W. Paufler, B. Böning, and S. Fritzsche, "High harmonic generation with Laguerre–Gaussian beams," *Journal of Optics*, vol. 21, no. 9. IOP Publishing, p. 094001, 25-Jul-2019.
[20] J. Arlt, K. Dholakia, L. Allen, and M. J. Padgett, "The production of multiringed Laguerre–Gaussian modes by computer-generated holograms," *Journal of Modern Optics*, vol. 45, no. 6. Informa UK Limited, pp. 1231–1237, Jun-1998.
[21] M. K. Karahroudi, B. Parmoon, M. Qasemi, A. Mobashery, and H. Saghafifar, "Generation of perfect optical vortices using a Bessel–Gaussian beam diffracted by curved fork grating," *Applied Optics*, vol. 56, no. 21. Optica Publishing Group, p. 5817, 12-Jul-2017.
[22] L. Stoyanov, S. Topuzoski, I. Stefanov, L. Janicijevic, and A. Dreischuh, "Far field diffraction of an optical vortex beam by a fork-shaped grating," *Optics Communications*, vol. 350. Elsevier BV, pp. 301–308, Sep-2015.
[23] L. Janicijevic and S. Topuzoski, "Fresnel and Fraunhofer diffraction of a Gaussian laser beam by fork-shaped gratings," *Journal of the Optical Society of America A*, vol. 25, no. 11. Optica Publishing Group, p. 2659, 10-Oct-2008
[24] S. Topuzoski and L. Janicijevic, "Fraunhofer diffraction of a Laguerre–Gaussian laser beam by fork-shaped grating," *Journal of Modern Optics*, vol. 58, no. 2. Informa UK Limited, pp. 138–145, 20-Jan-2011.
[25] S. Topuzoski and Lj. Janicijevic, "Conversion of high-order Laguerre–Gaussian beams into Bessel beams of increased, reduced or zeroth order by use of a helical axicon," *Optics Communications*, vol. 282, no. 17. Elsevier BV, pp. 3426–3432, Sep-2009.
[26] Y. Liu *et al.*, "Generation of perfect vortex and vector beams based on Pancharatnam-Berry phase elements," *Scientific Reports*, vol. 7, no. 1. Springer Science and Business Media LLC, 09-Mar-2017.

[27] P. Vaity and L. Rusch, "Perfect vortex beam: Fourier transformation of a Bessel beam," *Optics Letters*, vol. 40, no. 4. Optica Publishing Group, p. 597, 11-Feb-2015.

[28] M. Erhard, R. Fickler, M. Krenn, and A. Zeilinger, "Twisted photons: new quantum perspectives in high dimensions," *Light: Science & Applications*, vol. 7, no. 3. Springer Science and Business Media LLC, pp. 17146–17146, 17-Oct-2017.

[29] M. W. Beijersbergen, R. P. C. Coerwinkel, M. Kristensen, and J. P. Woerdman, "Helical-wavefront laser beams produced with a spiral phaseplate," *Optics Communications*, vol. 112, no. 5–6. Elsevier BV, pp. 321–327, Dec-1994

[30] V. V. Kotlyar, A. A. Kovalev, and A. P. Porfirev, "Astigmatic transforms of an optical vortex for measurement of its topological charge," *Applied Optics*, vol. 56, no. 14. Optica Publishing Group, p. 4095, 05-May-2017

[31] S. Topuzoski, "Fresnel and Fraunhofer diffraction of (*l*,*n*)th-mode Laguerre–Gaussian laser beam by a fork-shaped grating," *Journal of Modern Optics*, vol. 66, no. 14. Informa UK Limited, pp. 1514–1527, 11-Jul-2019.

[32] J. Pereiro-García, M. García-de-Blas, M. A. Geday, X. Quintana, and M. Caño-García, "Flat variable liquid crystal diffractive spiral axicon enabling perfect vortex beams generation," *Scientific Reports*, vol. 13, no. 1. Springer Science and Business Media LLC, 10-Feb-2023.

[33] J. H. McLeod, "The Axicon: A New Type of Optical Element," *Journal of the Optical Society of America*, vol. 44, no. 8. Optica Publishing Group, p. 592, 01-Aug-1954.

[34] A. S. Rao, "A conceptual review on Bessel beams," *Physica Scripta*, vol. 99, no. 6. IOP Publishing, p. 062007, 22-May-2024.

[35] Birrittella, R. J., Alsing, P. M., & Gerry, C. C. (2021, March). The parity operator: Applications in quantum metrology. *AVS Quantum Science*. American Vacuum Society.

[36] K. Y. Bliokh, F. J. Rodríguez-Fortuño, F. Nori, and A. V. Zayats, "Spin–orbit interactions of light," *Nature Photonics*, vol. 9, no. 12. Springer Science and Business Media LLC, pp. 796–808, 27-Nov-2015.

[37] S. Frankel, "Complete Approximate Solutions of the Equation x = tan x," *National Mathematics Magazine*, vol. 11, no. 4. JSTOR, p. 177, Jan-1937.

[38] M. Opper and C. Archambeau, "The Variational Gaussian Approximation Revisited," *Neural Computation*, vol. 21, no. 3. MIT Press - Journals, pp. 786–792, Mar-2009.

[39] Ghosh, Sanmitra, Paul J. Birrell and Daniela De Angelis. "Variational inference for nonlinear ordinary differential equations." *International Conference on Artificial Intelligence and Statistics*, 2021.

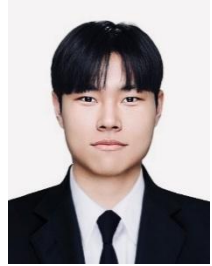

**First. Author. Junsu Kim[1]**

Department of Physics,
Incheon National University (INU),
Incheon, 22012, South Korea

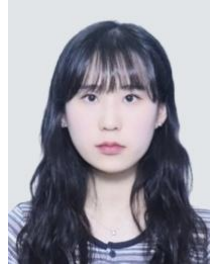

**Second. Author. Jiyeon Baek[2]**

Department of Information and Telecommunication Engineering,
Incheon National University (INU),
Incheon, 22012, South Korea

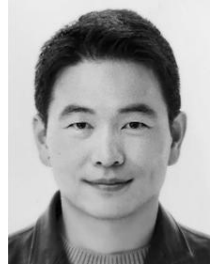

**Corresponding. Author. Hyunchae Chun[*]**

Department of Information and Telecommunication Engineering,
Incheon National University (INU),
Incheon, 22012, South Korea

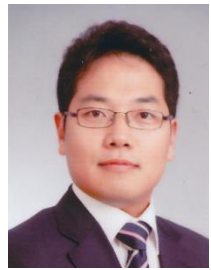

**Corresponding. Author. Seung Ryong Park[*]**

Department of Physics,
Incheon National University (INU),
Incheon, 22012, South Korea